\documentclass[aps,pra,longbibliography,superscriptaddress,twocolumn,10pt,nofootinbib]{revtex4-1}
\usepackage[utf8]{inputenc}
\usepackage[english]{babel}
\usepackage{times}
\usepackage{color}
\usepackage{graphicx}
\usepackage{import}
\usepackage{amsmath}
\usepackage{braket}
\usepackage{amssymb}
\usepackage{listings}
\usepackage{hyperref}
\usepackage[all]{hypcap}

\usepackage{booktabs}

\AtBeginDocument{
\heavyrulewidth=.08em
\lightrulewidth=.05em
\cmidrulewidth=.03em
\belowrulesep=.65ex
\belowbottomsep=0pt
\aboverulesep=.4ex
\abovetopsep=0pt
\cmidrulesep=\doublerulesep{}
\cmidrulekern=.5em
\defaultaddspace=.5em
}

\usepackage{tikz}
\usepackage{tikzscale}
\usetikzlibrary{backgrounds,math,fit,decorations.pathreplacing,calc,shapes}

\begin{document}
\title{A local and scalable lattice renormalization method for ballistic quantum computation}
\author{Daniel Herr}
\email{daniel.herr@riken.jp}
\affiliation{Quantum Condensed Matter Research Group, CEMS, RIKEN, Wako-shi, 351--0198, Japan}
\affiliation{Computational Physics, ETH Zurich, 8093 Zurich, Switzerland}
\author{Alexandru Paler}
\affiliation{Institute for Integrated Circuits, Johannes Kepler University, Linz, 4040, Austria}
\author{Simon J. Devitt}
\affiliation{ARC Centre for Engineered Quantum Systems, Department of Physics and Astronomy, Macquarie University, North Ryde, New South Wales 2109, Australia}
\affiliation{Turing Inc., 2601 Dana Street, Berkeley, CA 94704, USA.}
\author{Franco Nori}
\affiliation{Quantum Condensed Matter Research Group, CEMS, RIKEN, Wako-shi 351--0198, Japan}
\affiliation{Department of Physics, University of Michigan, Ann Arbor, MI 48109--1040, USA}

\begin{abstract}
 A recent proposal has shown that it is possible to perform linear-optics quantum computation using a ballistic generation of the lattice. Yet, due to the probabilistic generation of its cluster state, it is not possible to use the fault-tolerant Raussendorf lattice, which requires a lower failure rate during the entanglement-generation process. Previous work in this area showed proof-of-principle linear-optics quantum computation, while this paper presents an approach to it which is more practical, satisfying several key constraints. We develop a classical measurement scheme, that purifies a large faulty lattice to a smaller lattice with entanglement faults below threshold. A single application of this method can reduce the entanglement error rate to $7\%$ for an input failure rate of $25\%$. Thus, we can show that it is possible to achieve fault tolerance for ballistic methods.
\end{abstract}

\maketitle

\section{Introduction}
Several physical platforms are aiming at achieving quantum computing~\cite{buluta_review}. For example, a qubit can be implemented using superconductors~\cite{superconducting_err_cor}, silicon~\cite{VD14,TT16}, trapped ion systems~\cite{L16,LM16}, or using linear optics~\cite{Knill2001,kieling_quantumcomputation}. Major advances in the fidelity of these qubits have made the application of error-correction codes, such as the surface code~\cite{Fowler2012}, feasible. The surface code is of particular interest due to its high threshold~\cite{Fowler2012} and 2D nearest-neighbor layout.
While the surface code is suitable for qubit implementations which are relatively easy to control, linear-optics quantum computation~\cite{LOQC} is based on a slightly different principle, where photons are entangled in a cluster state which is then consumed during the computation.  This quantum one-way computer was proposed by Raussendorf et. al.~\cite{one_way_comp,RBB03}.
A high-level implementation for such a quantum computer can be divided into three steps~\cite{3steps01,3steps02}:
\begin{enumerate}
  \item Photon sources: delivers GHZ-triplets
  \item Entangling layer: generates the cluster state.
  \item Measurements: measurements in different bases allow for universal computation.
\end{enumerate}
While cluster states have also been studied for solid state qubits~\cite{cluster_nori1,cluster_nori2,cluster_nori3}, this scheme is better suited for photonics because it prioritizes measurements over the sequential application of quantum gates, and thus utilizes the ability to generate photons continuously.

The original proposal~\cite{one_way_comp,RBB03} was not protected against errors but a similar approach can introduce fault tolerance~\cite{raussendorf_lattice}. This approach uses the Raussendorf lattice as an underlying resource which protects the logical state both against noise and photon loss~\cite{Briegel2009}.

There are several approaches for creating a cluster state for quantum one-way computation~\cite{Briegel2009}. A drawback of linear optics is that the process of creating entanglement is non-deterministic. Thus, there remains a non-zero probability that each entanglement operation fails and the resulting lattice misses edges. Some approaches like~\cite{snowflake} try to remedy the probabilistic nature by adding redundancy to the entangling procedures. However, these approaches require many switches which rely on the outcome of previous entangling operations and add more noise to the system.
Another approach is to just use these non-deterministic gates and generate a faulty lattice. This ballistic approach~\cite{kieling_quantumcomputation} of linear-optics quantum computation recently gained attention due to improved theoretical entangling operations~\cite{bell_measure1,bell_measure2} which fail with $25\%$ probability. It has been pointed out that the error rate is still below the percolation threshold ($37.3\%$) calculated in~\cite{percolation_threshold}. Therefore, information can be transported from one end to the other given a large enough faulty lattice~\cite{scoop}.

In order to build a large scale quantum computer, this ballistic approach to linear-optics quantum computing should generate the Raussendorf lattice which then allows for fault-tolerant computation. Classical control software must now be developed in order to cope with $25\%$ of faulty entanglement operations, while still retaining these error-correction capabilities. In this paper, we provide an example of such an algorithm which is based on Ref.~\cite{scoop} and that acts as a preprocessing step. One should note that the approach in~\cite{scoop} has several limitations: ($i$) only in 1D; ($ii$) it is global; ($iii$) it is not fault-tolerant. Our approach has none of these drawbacks when it is combined with the usual Raussendorf lattice error-correction schemes~\cite{Raussendorf2007,photon_loss,entanglement_error}.
It should be noted that the preprocessing is not inherently fault-tolerant which results in a trade-off between the rate of missing bonds and an accumulation of errors due to imperfect measurements. The accumulation of these errors, however, only shifts the threshold of the Raussendorf lattice and can be remedied by higher fidelities of the experimental setup.
This is the first purification procedure for a 3D fault-tolerant lattice, with previous work in 1D being done in~\cite{scoop}. Thus, we expect future improvements to the algorithm using better heuristics that will relax the requirements for the experimental setup.

\section{Background}
In the following we will review relevant concepts. We will give the definition of graph states and explain how they can be modified. Then we will review the Raussendorf lattice, its creation, and its error-correction capabilities with a focus on faulty edges.

\subsection{Graph States}
\begin{figure}
  \centering
  \includegraphics[width=0.8\columnwidth]{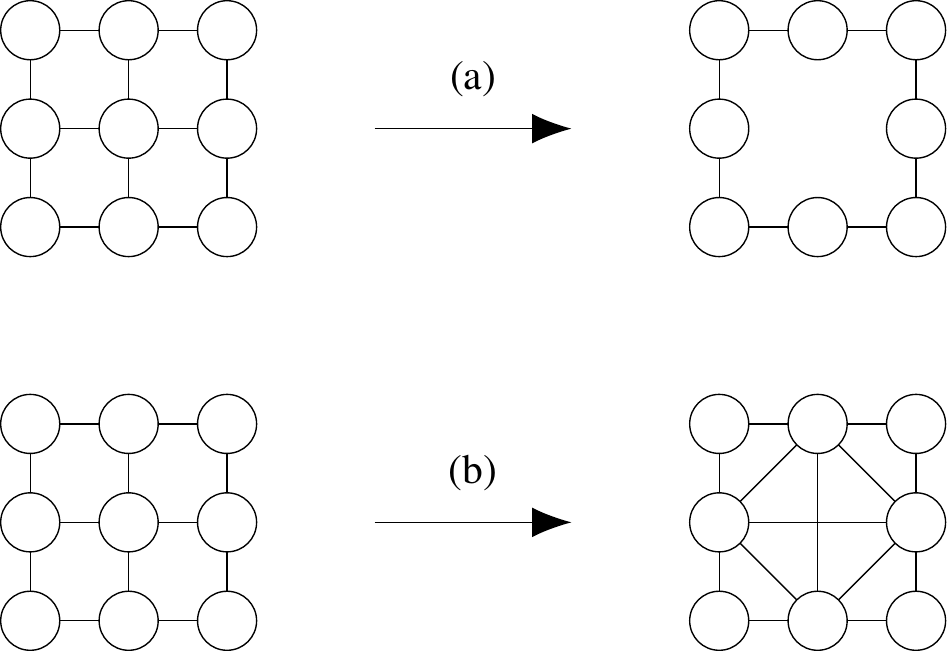}
  \caption{\label{fig:graphmanip_ex} The figure shows how a regular lattice changes under a $Z$-basis (a) or a $Y$-basis (b) measurement. These are the only measurement operation that our purification procedure will need.}
\end{figure}

Graph states~\cite{hein_graph_state} are a generalization of cluster states~\cite{cluster_states} and can be described using a undirected graph $G=\left(V,E\right)$ with vertices $V$ and edges $E$. Each vertex corresponds to a physical qubit initialized in the $\ket{+}$-state. On each edge, a controlled-phase gate is applied. This results in the final state of:
\begin{equation*}
  \ket{\psi}_\text{Graph} = \prod_{(i,j) \in E} \text{CZ}_{i,j} \ket{+}^{\otimes\left|V\right|}.
\end{equation*}
Using measurements, the graph can be modified to another graph. The modification rules have been discussed in Ref.~\cite{hein_graph_state}. Our proposal will only rely on two particular easy measurement operations:
\begin{enumerate}
  \item $Z$-measurement on qubit $a$: Remove $\left\{a\right\}$ from the graph and break all connections it was involved in.
  \item $Y$-measurement on qubit $a$: Remove $\left\{a\right\}$ and add connections between neighbors. This method can be used to generate long-distance edges.
\end{enumerate}
An example of these two rules on a square lattice is shown in Figure~\ref{fig:graphmanip_ex}.

\subsection{Creation of the lattice}
  \begin{figure}
    \centering
    \includegraphics[width=1\columnwidth]{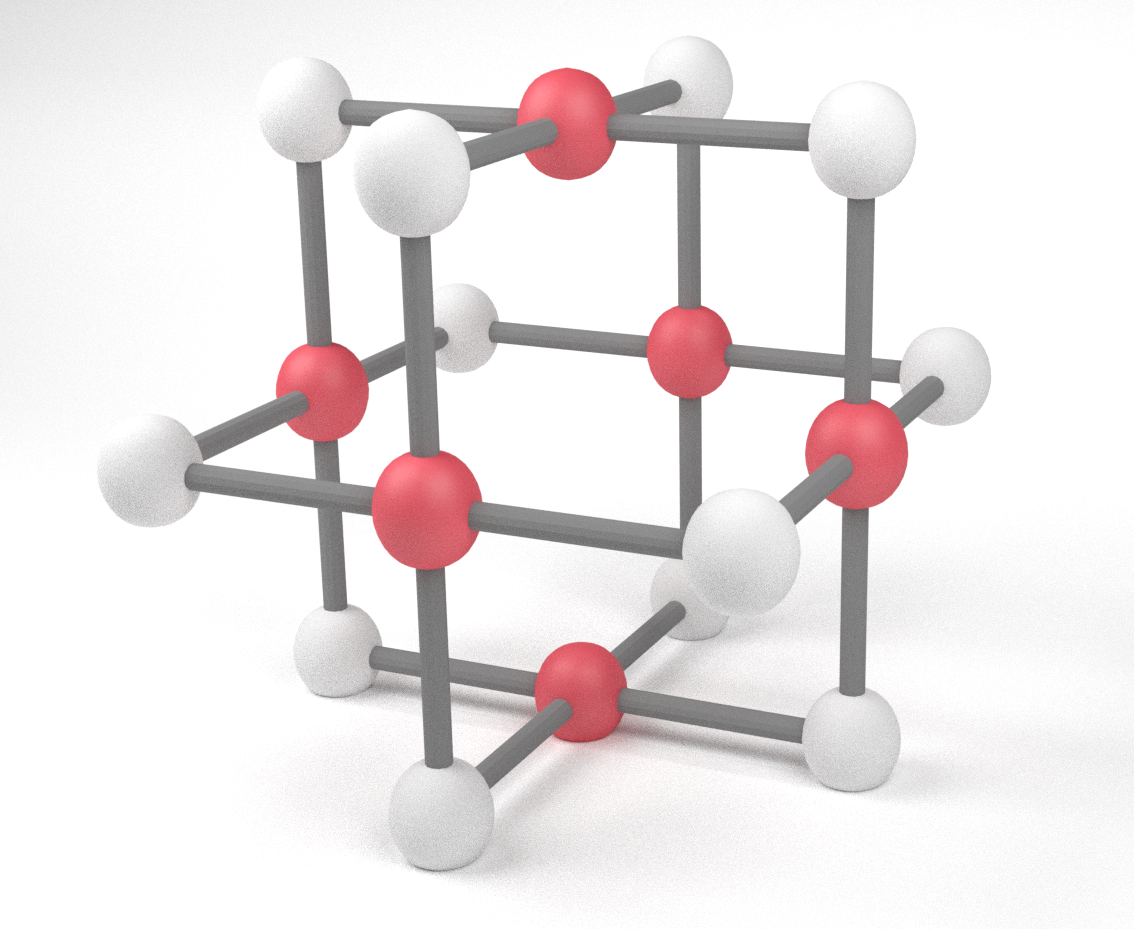}
    \caption{\label{fig:Raussendorf_unitcell} A unit cell of the Raussendorf lattice. The spheres represent individual photons and the connections between them represent entanglement given by the definition of a graph state. The photons which are colored in red contribute to a single $X$-parity check. Whereas the white spheres correspond to the faces of the dual lattice, which make up $Z$-parity check operations.}
  \end{figure}

A special graph state with error-correction capabilities is given by the Raussendorf lattice~\cite{raussendorf_lattice}. It is a 3D lattice whose unit cell is shown in figure~\ref{fig:Raussendorf_unitcell}.

\begin{figure}
    \centering
    \includegraphics[width=\columnwidth]{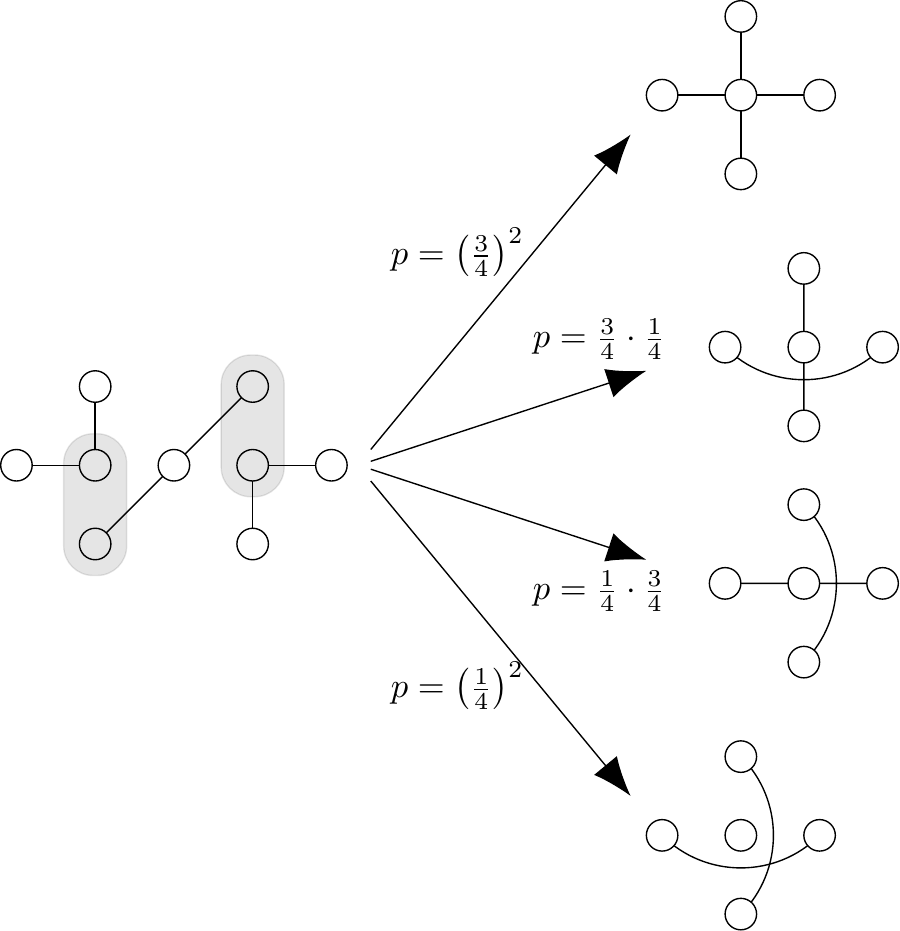}
    \caption{\label{fig:microcluster} This figure shows the creation of micro-clusters as described in~\cite{ballistic}. The areas shaded in grey indicate which two qubits are used for the fusion gates. Depending on the measurement outcome of these fusion gates, the structure will take one of the shapes on the right.}
  \end{figure}

To create the Raussendorf lattice in a ballistic way, GHZ states are needed as a resource. Three of these GHZ states can be entangled to a micro-cluster, using two probabilistic fusion gates~\cite{ballistic}. There are two ways to apply these fusion gates, with different additional resources. The fusion gate given in Ref.~\cite{bell_measure1} requires an additional pair of maximally entangled photons, whereas Ref.~\cite{bell_measure2} requires four single photons. The creation of the micro-clusters follows~\cite{ballistic} and all possible outcomes of the micro-cluster generation are shown in Figure~\ref{fig:microcluster}.

The central node of each micro-cluster will correspond to a node in the final lattice, while its surrounding nodes are consumed in additional fusion operations to connect clusters with each other. It can be seen in Figure~\ref{fig:microcluster} that a failure during the generation of these micro-clusters results in non-local entanglement. However, it becomes exponentially unlikely for edges with larger distances.

After the creation of the micro-clusters, each of these needs to be entangled to its neighbors on the large lattice. This is where our proposal deviates from~\cite{ballistic}, since the underlying lattice we try to implement is the Raussendorf lattice and not the diamond lattice from the original proposal.

\begin{figure}
    \centering
    \includegraphics[width=\columnwidth]{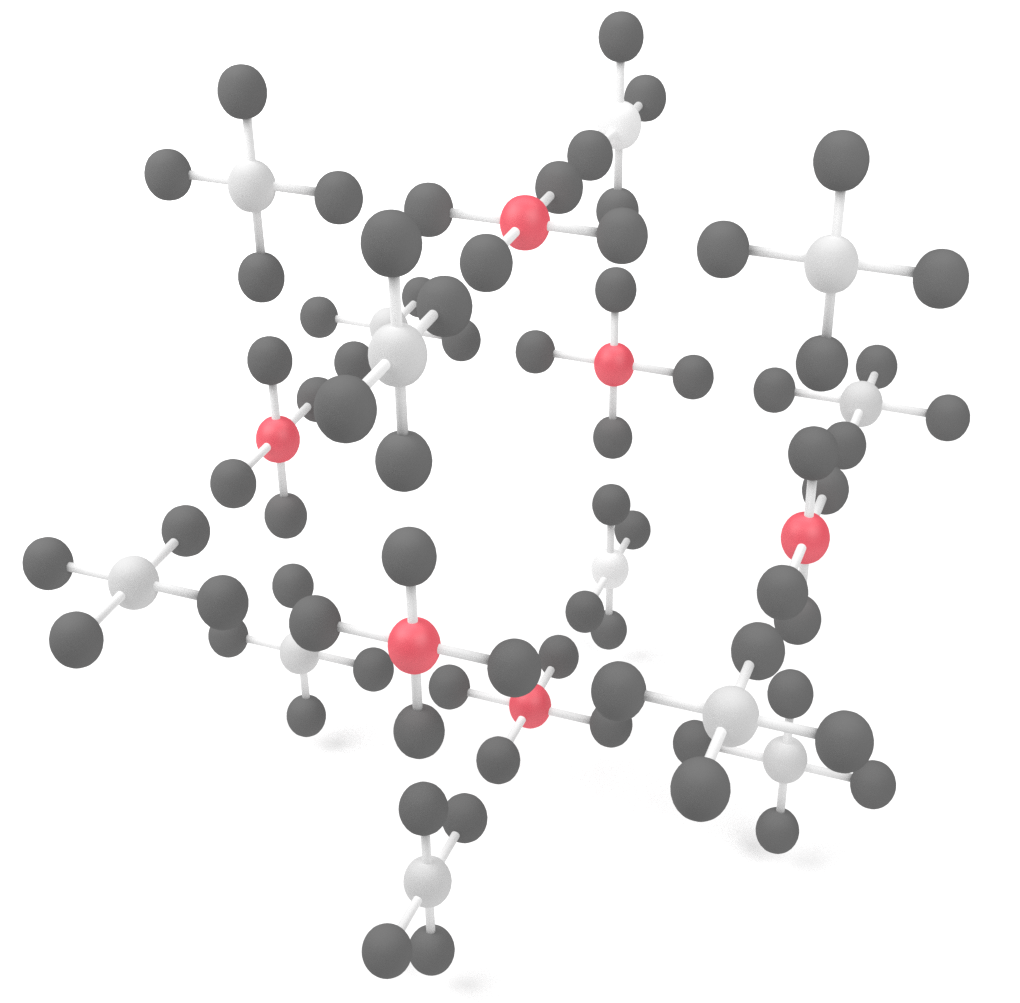}
    \caption{\label{fig:Raussendorf_gen} This figure shows how to build the Raussendorf unit cell using properly aligned micro-clusters. All neighboring pairs of dark nodes are consumed during the application of the fusion operations. The remaining white and red nodes correspond to the same color of the nodes from Figure~\ref{fig:Raussendorf_unitcell}. For simplicity, only successful micro-clusters are shown here. See Figure~\ref{fig:microcluster} to see all possible micro-cluster shapes.}
  \end{figure}

In Figure~\ref{fig:Raussendorf_gen} the generation of this lattice is shown. Each micro-cluster will correspond to a single node after all fusion gates have been performed. The fusion of these micro-clusters happens with a probability of $75\%$. Thus, $25\%$ of the time the creation of bonds in this lattice fails.

\subsection{Error correction}

Error detection and correction can be done using parity checks on particular nodes on the Raussendorf lattice. As an example, the total parity in the $X$-basis of the qubits colored in red (Figure~\ref{fig:Raussendorf_unitcell}) is conserved unless there has been an error. A self-similar lattice which is shifted by half a unit-cell uses the photons shown in white to perform $Z$-parity checks.

These parity checks together enable the error-correction capabilities of the Raussendorf lattice. Furthermore, with the method of Ref.~\cite{photon_loss} the Raussendorf lattice is also protected against photon losses. The main idea in this approach is to use the linearity of the parity checks to form super cells and perform parity checks on these. This resulted in a trade-off between the error rate due to faulty measurements and the rate of photon loss. The best photon loss rates that could still be corrected were around $25\%$~\cite{photon_loss}.

A lattice with faulty edges can be translated to a lattice with missing nodes, by deliberately losing one of the photons at the end of a faulty edge. This is done by performing a measurement in $Z$-basis on one of these photons. A recent paper~\cite{entanglement_error} described this as an adaptive correction scheme where the measurement basis needs to be changed depending on the error. This adaptive scheme can tolerate a loss rate of $14.5\%$ of all edges.
Another approach is to keep measuring as usual and then in the classical tracking software treat both qubits that are involved in the faulty connection as lost photons. There, still correctable loss rates lie at around $6.5\%$.
Unfortunately, neither approach can correct for error rates of $25\%$ and, thus, preprocessing in some form has to be performed.

\section{Graph purification: General Idea}
The general idea of our graph purification proposal is to develop a measurement scheme that translates a large Raussendorf lattice with many faults into a smaller Raussendorf lattice with fewer faults. Our procedure is based on Ref.~\cite{scoop} which investigated how path-finding procedures can help for quantum computation on a faulty lattice. It is not a quantum error correcting code, such that errors will accumulate during this step. Nevertheless, after this preprocessing, the original lattice has been translated to a lattice with {\it fewer\/} faults such that a general error-correction procedure can be used.

The main requirements for such an algorithm are
\begin{enumerate}
  \item The algorithm should be local, i.\ e. the algorithm's corrections should only rely on faults in the vicinity of the lattice. This is important since the lattice is generated continuously and only a part of it is physically available at any time.
  \item The algorithm should give the corrections fast. This is important, because photons are fast and delays in computation translate to large sizes of the quantum computer with long optical fibers.
  \item The algorithm should require as little overhead as possible in terms of photons.
  \item Scalability: adding more photons should be possible (e.g. the algorithm should be parallelizable).
\end{enumerate}

We will now present our scheme to purify the faulty lattice. It is based on the idea that while a $25\%$ error rate is very high, it is still below the percolation threshold of the Raussendorf lattice. For a larger lattice, the probability to find paths from one node to another increases.
Nodes from the large and faulty Raussendorf lattice are chosen and will make up the purified Raussendorf lattice. These nodes are then connected by finding paths through the faulty lattice. All photons on such paths have to be measured in the $Y$-basis and will therefore create edges between the chosen nodes. All other qubits are measured in the $Z$-basis and are thus removed from the lattice.

\section{Implementation}

\begin{figure}
  \centering
  \includegraphics[width=\columnwidth]{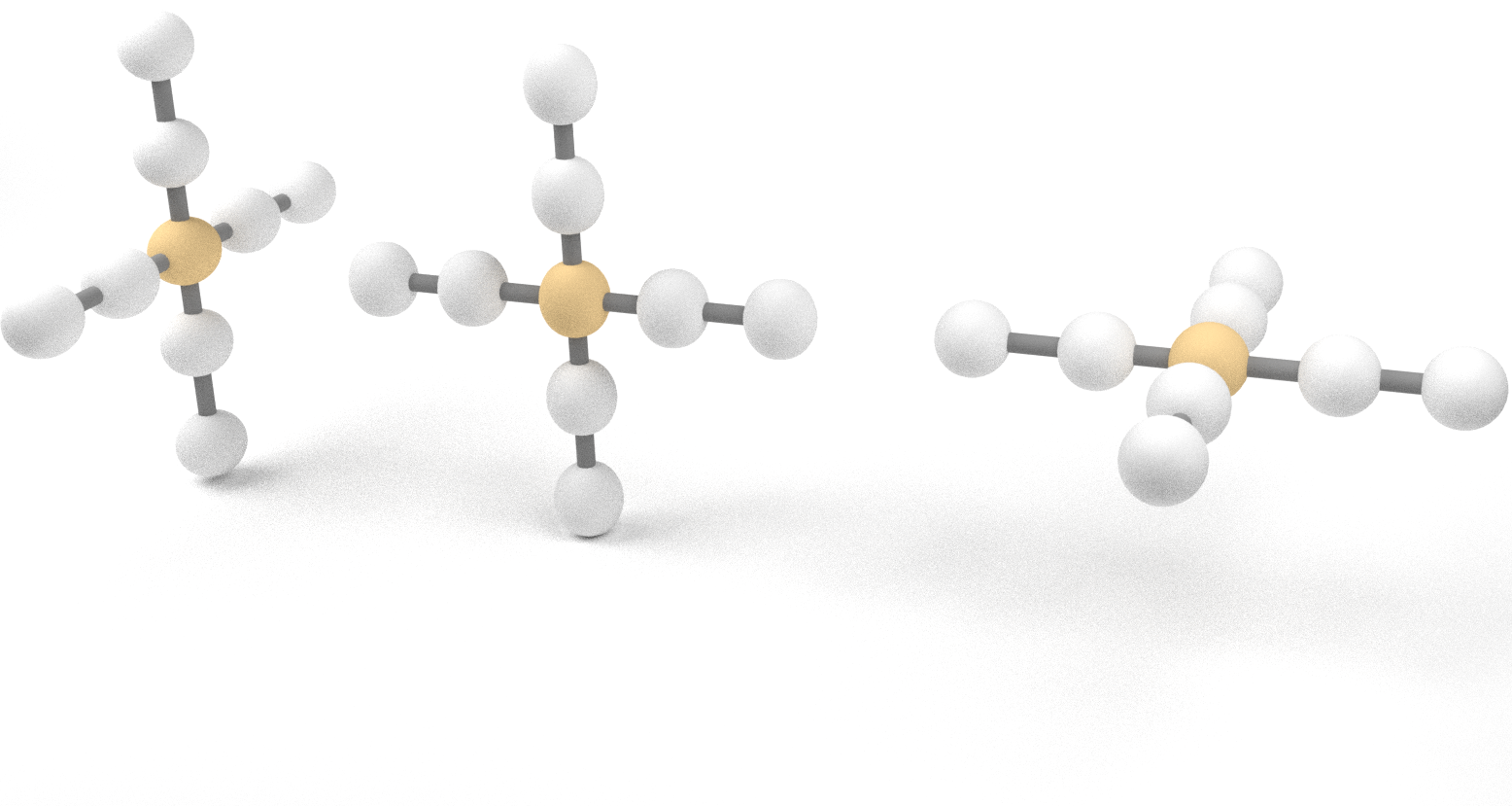}
  \caption{\label{fig:structures} Here, all possible orientations of a single structure are shown. The node in the center will be used as the node for the purified lattice.}
\end{figure}

The code to this description is open-source and hosted on Github~\cite{code}.

The implementation divides the original Raussendorf lattice into boxes. Each box has the same number of qubits along its edges. This size will be referred to as box size throughout the paper. In each box, one of the structures from Figure~\ref{fig:structures} is chosen. Each structure contains four handles which are the start or end points for the path-finding algorithm. To improve the performance of the algorithm, we added a heuristic method to select the best structure position: Of all possible structure positions the one with the highest number of neighbors from its handles is chosen. It should be noted that due to non-local connections, the neighbors of the structure are not necessarily nearest neighbors physically, but only neighbors due to the underlying graph.

After all the structures have been found, a path-finding algorithm needs to connect neighboring structures with each other. For example, our implementation uses the A* algorithm~\cite{Astar} with the Manhattan-norm as its heuristic function. This heuristic should give a decent estimate on the remaining distance but its distance estimate is not strictly smaller than the actual distance because of non-local entanglement due to the fusion gates. Therefore, A* is not guaranteed to find the shortest path, but for our algorithm finding the shortest paths this is not needed.

The code is written in C++ and was logically divided into three classes of which one implements the lattice of a single box, another class combines all boxes to the larger lattice, and the last class finds paths between different structures.

The lattice implementation for each box is given by the class \lstinline|Graph|. It implements the graph as a \lstinline|std::deque|, whose key is a unique identifier for the individual node and the value is a \lstinline|std::vector| of all neighbors. These neighbours are stored as a \lstinline|std::pair| where the first value is the id of the box and the second value gives the id of the node inside that box.
Further important functions are \lstinline|Graph::generate_connections| which randomly generates the lattice using the rules for the fusion operations and \lstinline|Graph::find_structure| which looks for a suitable position for the structure.

The large lattice class, \lstinline|Parallel|, contains a \lstinline|std::vector| of the class \lstinline|Graph|. This vector contains all the information related to the lattice. The class handles all high-level operations, such as output of the purified lattice, and calculations for the statistics. It further determines between which structures a path needs to be found. While our implementation is not yet parallel, the parallelization should be straightforward to implement in this class.

Finally, the class \lstinline|Astar|, implements the path-finding algorithm A*.

\section{Results}

\begin{figure}
  \centering
  \includegraphics[width=\columnwidth]{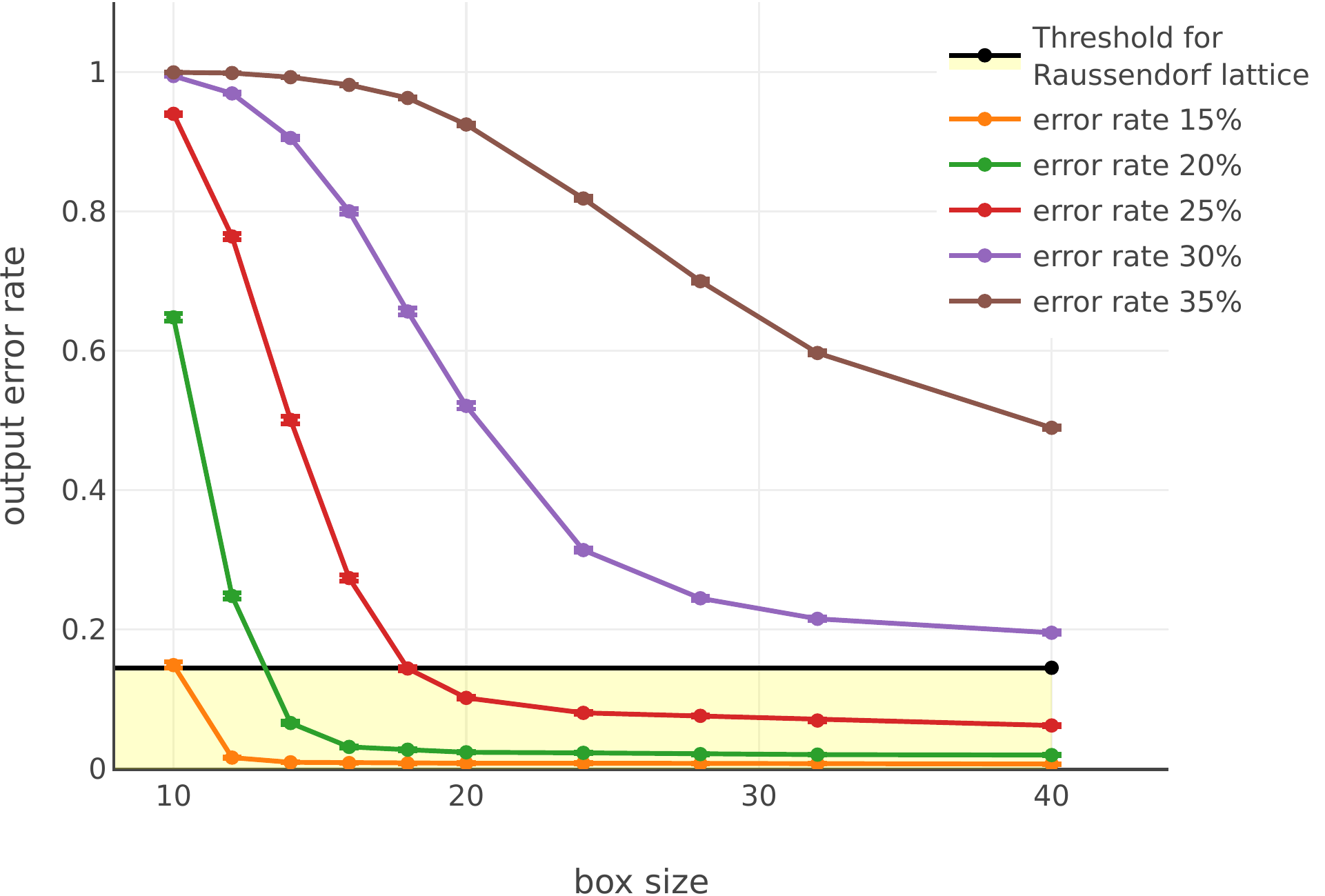}
  \caption{\label{fig:distance_plot} This plot shows the output error rate as a function of the box size. The black horizontal line indicates the threshold where the Raussendorf lattice can correct for missing bonds. For a fusion failure rate of $25\%$ the purified lattice for box sizes above $18$ are below the threshold of the Raussendorf lattice.}
\end{figure}

\begin{figure}
  \centering
  \includegraphics[width=1\columnwidth]{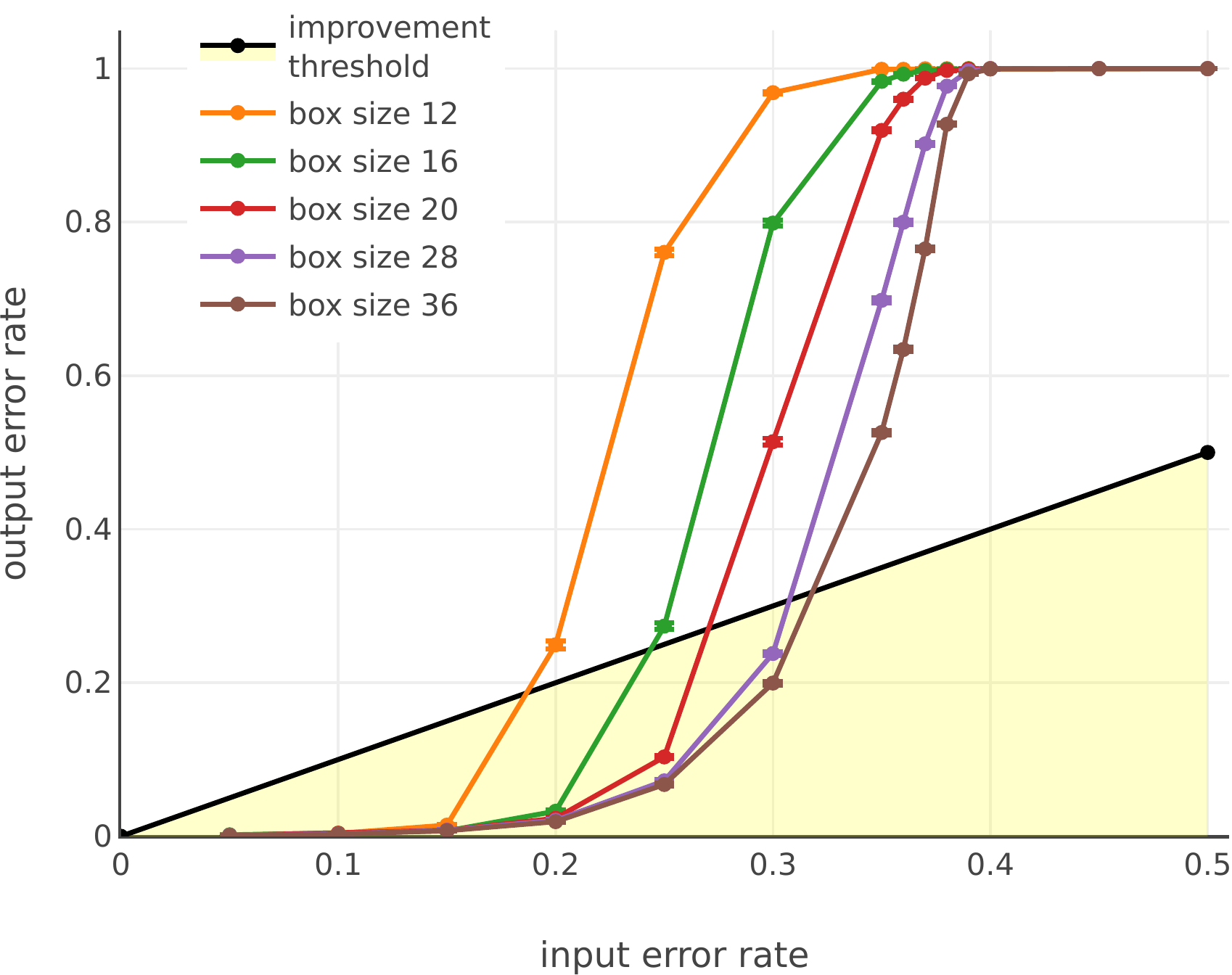}
  \caption{\label{fig:input_vs_output} This plot shows the output error rate as a function of the input error rate. The black line shows the border below which this purification algorithm can decrease the failure rate. Thus, it only makes sense to use this algorithm below an input error rate of around $32\%$ if box sizes up to 36 are used.}
\end{figure}
\begin{figure}
  \centering
  \includegraphics[width=1\columnwidth]{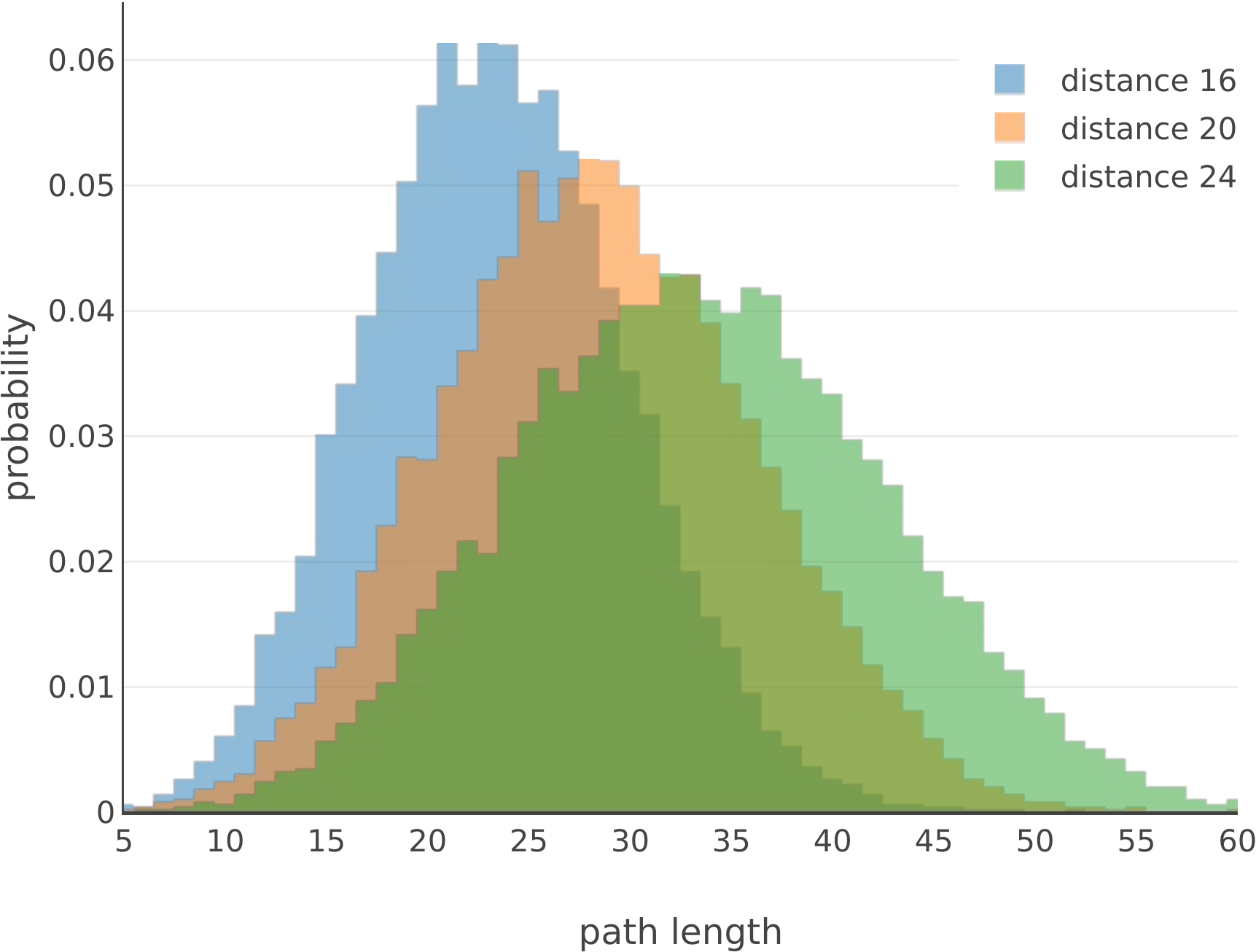}
  \caption{\label{fig:length_dist} This histogram shows the length distribution for all found paths. The mean lies at $23.48 \pm 0.04$ for box size 16, at $28.42 \pm 0.05$ for box size 20, and at $33.54 \pm 0.06$ for box size 24. All probabilities were calculated with at least $18,000$ different paths and using a constant fusion failure rate of $25\%$.}
\end{figure}

In order to see how well our code performs we ran this algorithm on lattices created with different success probabilities for the fusion gates, using different box sizes, and compared the rate of faults after our algorithm ran. The behavior of the output error rate with increasing size is plotted in Figure~\ref{fig:distance_plot}. One can see that for an initial failure rate of $25\%$ it is possible to reach an output error rate of about $7\%$ for the purified lattice. This is below the threshold rate of $14.5\%$~\cite{entanglement_error} of the Raussendorf lattice. Thus, it should be possible to use this code as a preprocessor for fault-tolerant ballistic quantum computation.

In Figure~\ref{fig:input_vs_output} the relation between input error rate and output error rate is shown. Every data point below the black curve shows an improvement over the input error rate. Thus, it makes sense to use this algorithm for input error rates below $32\%$.

Figure~\ref{fig:length_dist} shows a histogram of the length distribution of the paths. The average path length is larger than the box size because the shortest possible path is not always possible due to missing edges on the graph. It is possible to obtain shorter paths due to non-local interactions and differences in structure positions. The average path length for a box size of 20 is given by $28.42 \pm 0.05$. We will use these values in the following analysis to estimate the effects of errors.

\begin{figure}
  \centering
  \includegraphics[width=1\columnwidth]{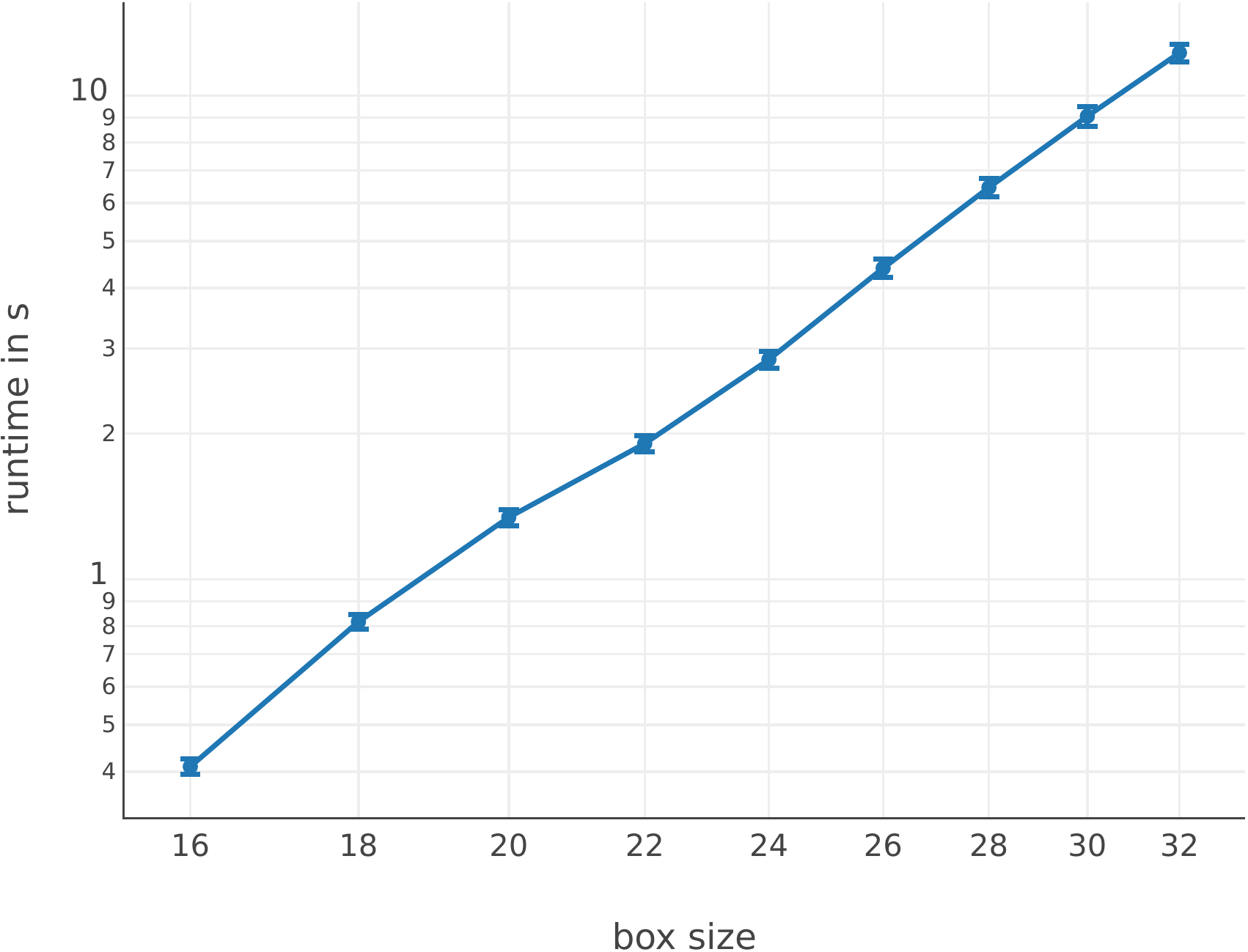}
  \caption{\label{fig:timing} This figure shows the absolute time that the algorithm needs to find structures and paths (the generation of the lattice was not measured). The plot has been performed for a constant number of boxes of $5\times5\times3$ and at a constant error rate for the fusion gates of $25\%$.}
\end{figure}

We performed a simple timing analysis by running the algorithm on a single core of an i7-4558U (2.8 GHz) CPU, to give a rough estimate on the speed of the algorithm. The results are plotted in Figure~\ref{fig:timing}. For a box size of $20$ and a $5\times5\times3$ lattice of boxes, the algorithm needs on average $1.34 \pm 0.05 \text{ s}$. However, it should be noted that not much effort was put into optimization and better performance can be expected from optimized implementations. The scaling of this algorithm is polynomially both in box size and number of boxes.

\section{Drawbacks of this Method}
\begin{figure}
  \centering
  \includegraphics[width=.8\columnwidth]{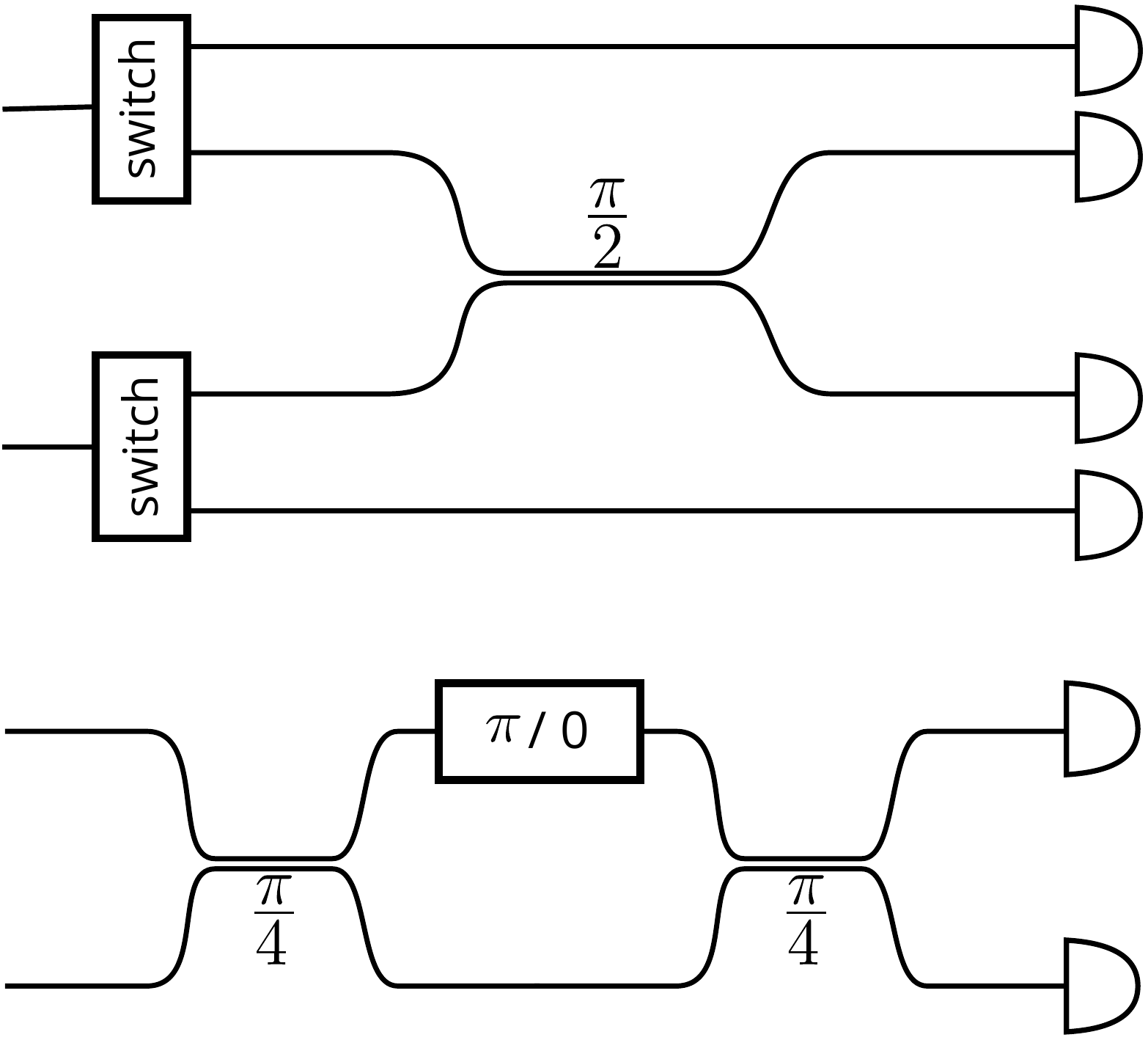}
  \caption{\label{fig:measurement_setup} This schematic shows two possible procedures for the measurements~\cite{OBrien}.}
\end{figure}

The purification process is not inherently fault-tolerant so errors can accumulate. In the following, we want to discuss the effects on the error rate that our approach has.

To analyse these sources of error it makes sense to discuss the measurement procedure. A qubit state is encoded as a spatial mode, which can be measured using a pair of photon detectors.
There are two ways of how a Hadamard operation can be implemented~\ref{fig:measurement_setup} to enable a change in measurement basis. Because our proposal requires to change the measurement basis depending on the lattice that has been created, the ability to add a Hadamard operator in reasonably short time has to be guaranteed. A simple Hadamard operation can be implemented by bringing together the two wave-guides for a length of $\pi / 2$. In order to add a choice of measurement basis, one can use switches as shown in the first approach of Figure~\ref{fig:measurement_setup}.
The other approach divides the Hadamard operation into two by bringing together both wave-guides for a length of $\pi / 4$ both times and adds a gate that creates a phase-difference between the two wave-guides. If this phase difference is zero, the Hadamard operation is performed, but if the phase-difference is $\pi$ one will obtain the identity operation.

There are several sources of errors that can occur in such setups. One source of errors comes from imperfections due to switches which will lead to photon loss. Another source of errors are imperfect rotations and Hadamard operations. These will lead to a shifted measurement basis and a small rotation in the state, which is teleported to the next qubit. The last source of errors is at the detectors. They could show false positives (the detector detects a non-existent photon) as well as false negatives (a photon is not registered at the detector).

For linear-optics applications the implementation of Hadamard operations is precise~\cite{OBrien,Obrien2} and using the second approach of Figure~\ref{fig:measurement_setup}, switches are not needed during measurement. Thus, in our analysis we will neglect both of these types of errors and concentrate on errors at the detectors.

All first-order errors at the detectors result in a nonsensical measurement: Both photon detectors are triggered at the same time or neither of the photon detectors is triggered. If such a case happens, it is clear that an error occurred but it is impossible to know the nature of the error. A second order error will result in the opposite measurement outcome. Obtaining the wrong measurement error will result in a Pauli error because, in measurement-based quantum computation, a by-product Pauli operation has to be applied depending on the measurement outcome.

One possible correction scheme for first-order errors is to choose a measurement outcome randomly. With $50\%$ this measurement is incorrect and due to the rules of measurement-based quantum computing a wrong by-product Pauli operator is applied. In the end, additional Pauli errors appear on the purified lattice and the Raussendorf lattice has to locate and identify them.

The total error rate for each path can be calculated using:
\begin{equation*}
P_\text{err} = 1 - f^{2\overline{L}}.
\end{equation*}
At box size $20$, the mean length of a path is $\overline{L}=29$ and given a detector fidelity $f=0.9999$ the resulting error rate is $P_\text{err}=0.57\%$ for each bond. The factor 2 in the equation comes from the fact that each measurement involves 2 photon detectors.

To obtain the error rate per node we assume that if a measurement error occurs we attribute it to the node in the same box. For a single node there are on average $4 \left(\overline{L} /2 \right)$ qubits for all 4 paths whose length in a single box corresponds to $\overline{L}/2$ each. Using this in the exponent the resulting error rate on each node is $P_\text{NodeErr}=1.15\%$. However, due randomly applying one of two correctional gates for first-order errors this error rate can be halved. The effective error rate per node is $P_\text{NodeErr} \approx 0.58\%$.

At box size $20$, the rate of failed connections is $10\%$. To be below threshold the remaining measurement errors need to be below $0.6\%$, which can be achieved with a fidelity of about $f=0.9999$.
This is a very strong requirement for the measurement setup but with improvements in the preprocessing algorithm it can be relaxed.

\section{Possible Improvements}
The algorithm seems to depend heavily on the type of structures and their position. We already used a heuristic that maximizes the possibilities for the first step of the path-finding algorithm but more advanced heuristics might improve the error rate of the purified lattice even further. Furthermore, a clustering algorithm should help in choosing good structure positions at the cost of an increased runtime. The effects of this should be included in future analysis.
Changes to the distance heuristic for the A*-search might also affect the performance of this proposal, but it was not investigated here.

\begin{figure}
  \centering
  \includegraphics[width=\columnwidth]{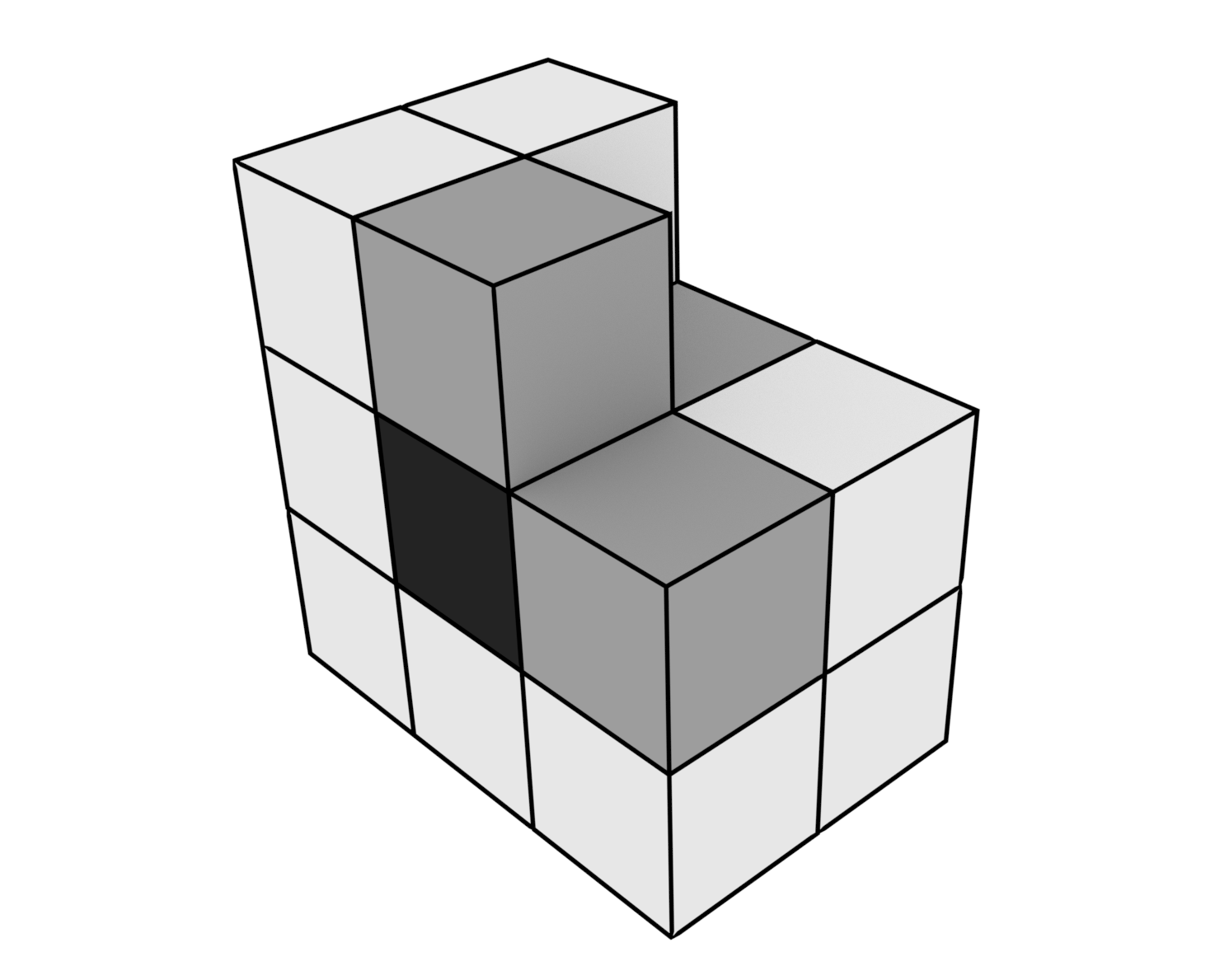}
  \caption{\label{fig:parallel1} This shows how to parallelize the algorithm for a big lattice. The process responsible for the calculation of the black box only requires information about the three nearest-neighbour grey boxes. Only this information needs to be exchanged.}
\end{figure}

This method is easily parallelize-able. Each processor could have its own set of boxes. Only information about the direct neighboring boxes needs to be exchanged with other processors. In Figure~\ref{fig:parallel1} the black box needs information about the boxes colored in gray only.
Every process needs to find a structure position in each of its boxes. Each process needs to send the position of its qubit-structures which lie on the boundary surface to the process on the left, and down (opposite direction). The box in the back will be treated by the same process so no communication is required. After every box received the information of its two neighbors, it can continue to find three paths in the right, up, and back directions.
The overhead of communication scales with the surface and not the volume and each process only needs to know a small part of the whole lattice, such that memory problems can be avoided.

\section{Workflow}

\begin{figure}
  \centering
  \includegraphics[width=\columnwidth]{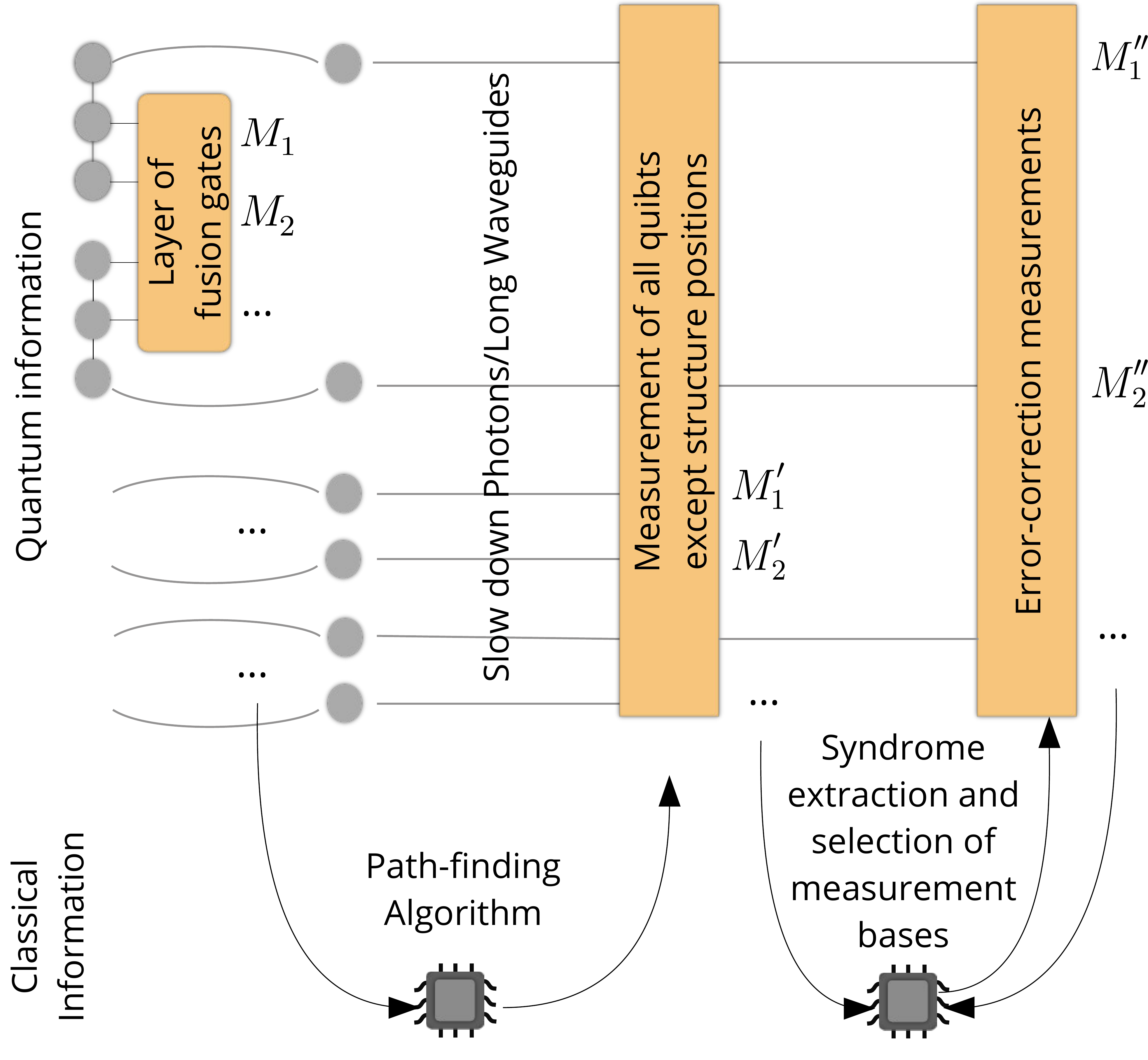}
  \caption{\label{fig:information_flow} This figure shows the quantum and classical flow of information. Starting from GHZ-states, some qubits will be measured in a layer of fusion gates and the measurement information $M_i$ is fed to the classical path-finding algorithm. Meanwhile, the remaining photons are led along a long wave-guide to give enough time to the classical computation. These qubits are then fed into a layer of measurements, where all photons except the single photons at the center of each structure are measured out. The measurements ${M'}_{i}$ are used to determine if any errors happened. This information is then sent to the Raussendorf error-correction processor that takes care of the syndrome extractions and measurements in proper bases.}
\end{figure}

In the introduction we mentioned a high-level design of a linear-optics quantum computer. Here, we want to refine on it, with the inclusion of our purification step. To this end we show in Figure~\ref{fig:information_flow} a possible quantum and classical flow of information and the actions that are taken due to that information.
To the left of the figure three-qubit GHZ-states are created and used by fusion gates. It should be noted that additional GHZ states and photons are needed in that step but for simplicity they are not shown here. Using the measurement results of the fusion gates, the classical path-finding algorithm can infer the graph state that was generated and find paths connecting nearest-neighbor structures. All photons are measured out except for the center nodes of each structure. The routing of these center nodes requires a switch each and their measurement can be adjusted due to measurement errors on the path.
Finally, the last measurements are the actual measurements needed for the Raussendorf lattice, where syndrome extraction and the actual fault-tolerant quantum computation happen.

\section{Conclusion}
We have presented a way to purify the 3D lattice obtained from the ballistic procedure proposed in~\cite{ballistic} using ideas from~\cite{scoop}. This purification process can suppress entanglement errors due to probabilistic fusion gates and bring the error rate from $25\%$ down below the Raussendorf lattice threshold. This procedure, however, has the cost that errors along generated paths can accumulate and requires higher precision in measurement operations. Nevertheless, this approach shows that fault-tolerant quantum computation using ballistic lattice generation is possible. Looking back at the requirements we posed, we can see that our proposal fulfils several of them:
\begin{enumerate}
  \item The algorithm is local. Due to the exponential decay in large distance edges all connected nodes are located in the same box or neighbouring boxes.
  \item The algorithm scales polynomially in lattice size, but our implementation should still be improved in terms of absolute speed.
  \item The overhead in terms of qubits could be better: each box consists of about $20^3$ nodes, which are all consumed to generate one node in the purified lattice. Errors also accumulate, with larger sizes.
  \item The algorithm is easily scalable, with only little communication required by different processes.
\end{enumerate}

While our code works, many improvements can be made to this preprocessing step, such as using different measurement schemes to create entanglement with $X$-basis measurements. Thus, it is likely that the output error rate and therefore resource requirements are further reduced.
Then, fair comparisons between different ways to generate the lattice such as~\cite{snowflake} and the ballistic approach with preprocessing should be made in terms of overhead for the Raussendorf lattice. 

\begin{acknowledgments}
D.H. is supported by the RIKEN IPA program and A.P. was supported by the Linz Institute of Technology project CHARON, grant number LITD13361001.
S.J.D. acknowledges support from the JSPS Grant-in-aid for Challenging Exploratory Research and from the Australian Research Council Centre of Excellence in Engineered Quantum Systems EQUS (Project CE110001013).
FN was partially supported by the 
IMPACT program of JST, the CREST grant No. JPMJCR1676,
MURI Center for Dynamic Magneto-Optics via the AFOSR Award No. FA9550-14-1-0040, 
the Japan Society for the Promotion of Science (KAKENHI), 
JSPS-RFBR grant No 17-52-50023, 
RIKEN-AIST Challenge Research Fund, 
and the Sir John Templeton Foundation.
\end{acknowledgments}
\bibliography{biblio}
\end{document}